\documentclass[conference]{IEEEtran}
\IEEEoverridecommandlockouts
\usepackage{cite}
\usepackage{amsmath,amssymb,amsfonts}
\usepackage{algorithmic}
\usepackage{graphicx}
\usepackage{textcomp}
\usepackage{xcolor}
\def\BibTeX{{\rm B\kern-.05em{\sc i\kern-.025em b}\kern-.08em
    T\kern-.1667em\lower.7ex\hbox{E}\kern-.125emX}}

\title{Antenna Coding Optimization for Pixel Antenna Empowered Wireless Communication Using Deep Learning with Heterogeneous Multi-Head Selection
\thanks{$^\dagger$ Corresponding author.}}

\author{ \IEEEauthorblockN{Binzhou Zuo$^{1}$, Shanpu Shen$^{2}$ and Hongyu Li$^{1\dagger}$}

\IEEEauthorblockA{$^{1}$ Internet of Things Thrust, The Hong Kong University of Science and Technology (Guangzhou), Guangzhou, China \\ E-mail: \texttt{binzhouzuo@hkust-gz.edu.cn} and \texttt{hongyuli@hkust-gz.edu.cn} }

\IEEEauthorblockA{$^{2}$ State Key Laboratory of Internet of Things for Smart City, University of Macau, Macau, China \\ E-mail: \texttt{shanpushen@um.edu.mo}  }
}

\begin{document}
\maketitle

\begin{abstract}
Pixel antenna is a promising antenna technology that enables flexible adjustment of radiation characteristics and enhancement of wireless systems through antenna coding. This work proposes a novel deep learning-based antenna coding optimization algorithm. Specifically, the proposed algorithm is supported by a heterogeneous multi-head selection mechanism, whose main idea is to train multiple neural networks based on various coding schemes and select the one that leads to the best system performance. 
% The proposed algorithm employs multiple neural networks to optimize the antenna coding under known channel knowledge. 
Unlike traditional heuristic searching-based algorithms that require high computational complexity to achieve satisfactory performance, the proposed data-driven deep learning approach can achieve 98\% of the performance achieved by the searching-based algorithms with significantly reduced computational complexity. Results demonstrate that in pixel antenna empowered single-input single-output systems, the proposed algorithm achieves a computational speed 81 times faster than the searching-based algorithm. For more complex pixel antenna empowered multiple-input multiple-output systems, the computational speed is 297 times faster than the existing searching-based algorithm. Benefiting from the high performance and low computational complexity, this algorithm demonstrates the significant potential of pixel antennas as a novel and practical technology to enhance wireless systems.
\end{abstract}

\begin{IEEEkeywords}
    Antenna coding, deep learning, heterogeneous multi-head selection, neural networks, pixel antennas. 
\end{IEEEkeywords}

\maketitle

% 在关键词之后，Introduction之前放置资助信息

\section{Introduction}

Antennas are pivotal components in wireless communication systems. The multiple-input multiple-output (MIMO) technology enabled by multiple antennas has provided milestone advancements from 3G to 5G and is expected to remain dominant in the upcoming 6G networks \cite{yang2015fifty}. While increasing the scale of MIMO is able to reach higher requirements in throughput, latency, and coverage for 6G networks, it may not always be a good choice due to the significantly increased complexity of signal processing, hardware complexity, and cost \cite{10379539}. 
Instead, exploring the new degrees of freedom provided by antennas themselves, by reconfiguring the antenna characteristics, is envisioned to enhance the 6G communication network.

Antennas in conventional MIMO systems typically have fixed configuration and characteristics, such as operating frequency, radiation pattern, and polarization, meaning that antennas are excluded from signal processing and system optimization, yielding limited performance. To overcome this limitation, pixel antennas, as a reconfigurable antenna technology, are a promising solution. Pixel antennas are constructed by discretizing the radiation surface into a small grid of pixels and interconnecting them with RF switches \cite{3zhang2022low}. 
By controlling the states of RF switches, the antenna topology and configuration can be adjusted to reconfigure the characteristics \cite{song2013efficient,zhang2022highly,zhang2020polarization}. 
% In this way, pixel antennas break through the fixed configuration of conventional antennas and provide additional degrees of freedom in wave manipulation \cite{4shen2025antenna}. 
Related to pixel antennas, there is an emerging technology called fluid antenna systems (FAS) \cite{5wong2020performance}, where the antenna can switch position within a linear region. This position-switchable capability enables FAS to adapt to channel by adjusting the position, so as to enhance performance in various applications, such as reducing outage probability \cite{6zhu2023modeling} and enhancing multiple access \cite{7new2023information}. 
Pixel antennas have been adopted to mimic the antenna movement of FAS by controlling states of RF switches \cite{9zhang2024pixel}, which preliminarily reveals the potential of pixel antennas in wireless systems.
% An interesting observation made in \cite{9zhang2024pixel} is that pixel antennas can be used to implement FAS, by controlling the states of switches to mimic the antenna movement. This fact, together with existing applications of FAS \cite{10wang2024ai}, primarily reveals the potential to apply pixel antennas in wireless systems.

To further leverage the advantages of pixel antennas, a promising technique called antenna coding has been recently proposed for pixel antenna empowered wireless systems \cite{4shen2025antenna}. 
In antenna coding, the RF switch states are represented by binary variables called antenna coder, which can be optimized to adapt to the channel and enhance the wireless system.
To optimize antenna coding for pixel antenna empowered systems, heuristic searching-based algorithms are utilized \cite{4shen2025antenna,1211143412}, which rely on exhaustive searching over a large range. Therefore, it poses a critical challenge to balance the performance and computational complexity. 

To overcome this challenge, in this work, we propose a deep learning-based antenna coding optimization algorithm using a heterogeneous multi-head selection mechanism (HMSM). 
Deep learning \cite{13van2023ai}, as a data-driven approach, is renowned for efficient computation and high accuracy in complex problems, offering a promising method to achieve a better performance-complexity tradeoff.
Specifically, the proposed algorithm with HMSM is inspired by the multi-head attention mechanism for dealing with highly nonlinear problems with enhanced accuracy and robustness \cite{14li2021diversity}, and built on the pixel antenna communication model \cite{4shen2025antenna}. Another innovation of the proposed algorithm is that it utilizes the binary/Gray coding scheme to compress binary coders into an antenna map in compact decimal format, drastically reducing computational complexity. 
% Within this framework, each attention head acts as a dedicated nonlinear mapper, approximating the discontinuous properties of the channel from different feature perspectives. 
% This design effectively decomposes the nonlinear fitting task due to the nonlinear pixel antenna modeling into multiple parallel subtasks, thereby distributing the modeling difficulty and enhancing both the accuracy and robustness of the model \cite{15vaswani2017attention}. 
The efficiency and effectiveness of the proposed algorithm is validated by considering a channel gain maximization in pixel antenna empowered single-input single-output (SISO) systems, followed by the extension to MIMO capacity maximization. Results show that the proposed algorithm achieves 98\% of the performance achieved by the heuristic searching-based algorithm, while significantly reducing the computational complexity by up to 297 times, demonstrating the benefits of the deep learning-based proposed algorithm. 

\textbf{Notations:} Boldface lower-case and upper-case letters indicate column vectors and matrices, respectively. $\mathbb{C}$ and $\mathbb{R}^+$ denote the set of complex and positive real numbers, respectively. $(\cdot)^*$, $(\cdot)^{\mathsf{T}}$, $(\cdot)^{\mathsf{H}}$, and $(\cdot)^{-1}$ denote the conjugate, transpose, conjugate-transpose, and inversion operations, respectively.
$\|\cdot\|$ denotes the $\ell$-norm of a vector.
$|\mathbf{A}|$ denotes the determinant of a matrix $\mathbf{A}$ and $|a|$ denotes the absolute value of a scalar $a$. 
$\Re\{\cdot\}$ and $\Im\{\cdot\}$ denote the real and imaginary parts of a matrix, vector, or scalar. 

\section{Antenna Coding Based on Pixel Antenna}

In this section, we introduce the antenna coding technique based on pixel antennas, including the model and optimization formulation.

\subsection{Pixel Antenna Model}
\begin{figure}
    \centering
    \includegraphics[width=0.95\linewidth]{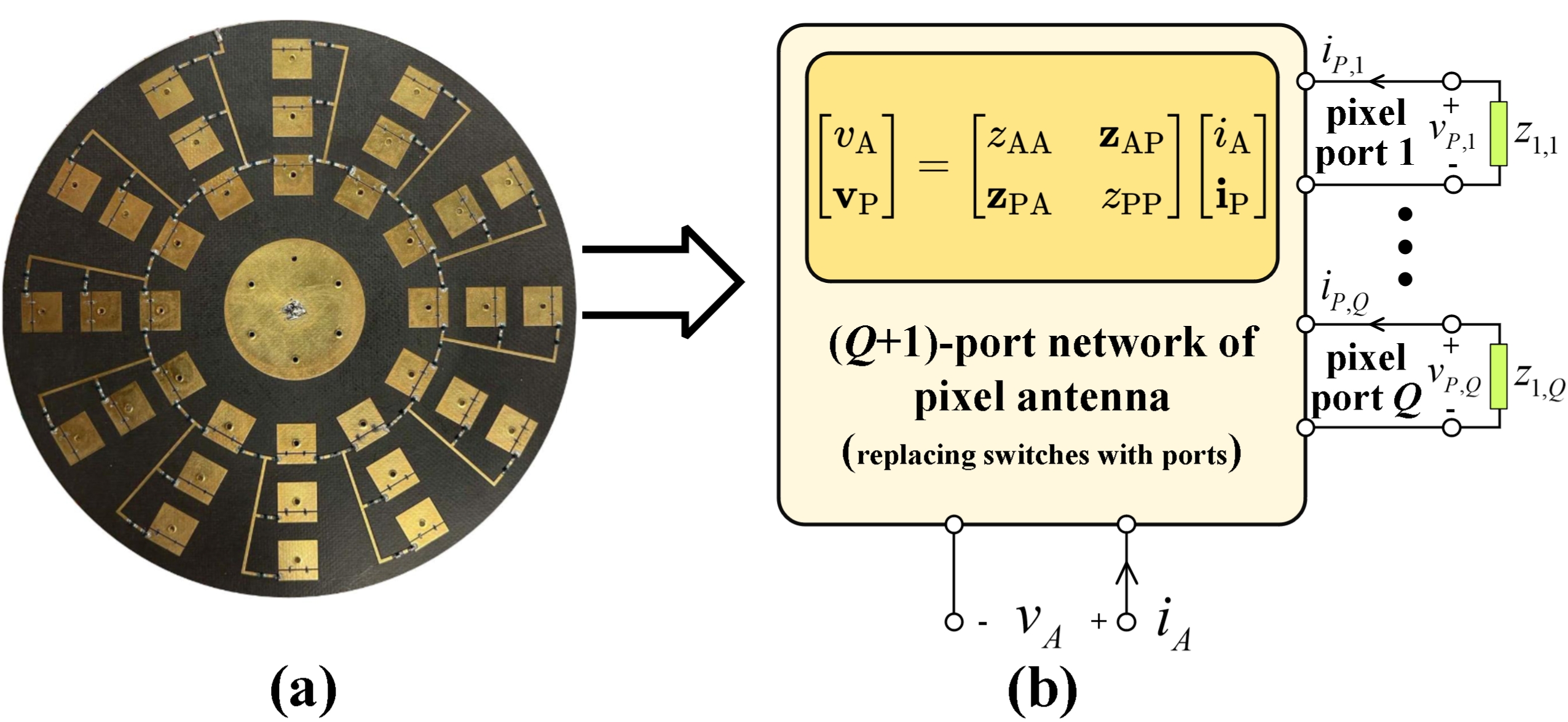}
    \caption{(a) An example of pixel antenna designs proposed in \cite{3zhang2022low} and (b) the multi-port circuit network for pixel antenna.}
    \label{fig:pixel_antenna_model}
\end{figure}

Pixel antennas are a reconfigurable antenna technology based on a discretized radiating surface embedded with RF switches. As shown in Fig. \ref{fig:pixel_antenna_model}(a), its core principle involves pixelizing a continuous radiating surface into numerous sub-wavelength units called pixels. Adjacent pixels are interconnected via RF switches \cite{3zhang2022low}. By controlling the on and off states of these switches, the antenna topology can be dynamically reconfigured, enabling real-time control over radiation characteristics such as radiation pattern.

Pixel antennas can be modeled using microwave multi-port network theory, as illustrated in Fig. \ref{fig:pixel_antenna_model}(b). A pixel antenna having $Q$ switches is represented as a $(Q+1)$-port network, comprising one antenna port and $Q$ pixel ports to model the $Q$ switches. This yields the following expressions for the antenna port voltage $v_\mathrm{A}\in\mathbb{C}$ and current $i_\mathrm{A}\in\mathbb{C}$, and the pixel port voltage vector $\mathbf{v}_\mathrm{P}\in\mathbb{C}^{Q\times 1}$ and current vector $\mathbf{i}_\mathrm{P}\in\mathbb{C}^{Q\times 1}$ as
\begin{equation}
    \begin{bmatrix}
    v_{\mathrm{A}} \\
    \mathbf{v}_{\mathrm{P}}
    \end{bmatrix} =
    \begin{bmatrix} z_\mathrm{AA} &\mathbf{z}_\mathrm{AP}\\
    \mathbf{z}_\mathrm{PA} & z_\mathrm{PP}\end{bmatrix}
    \begin{bmatrix}
    i_{\mathrm{A}} \\
    \mathbf{i}_{\mathrm{P}}
    \end{bmatrix},
\end{equation}
where 
% $\mathbf{Z}$ can be partitioned as $\mathbf{Z} = [z_\mathrm{AA},\mathbf{z}_\mathrm{PA}^\mathsf{T};\mathbf{z}_\mathrm{PA},\mathbf{Z}_\mathrm{PP}]$, with 
$z_\mathrm{AA}\in\mathbb{C}$ and $\mathbf{Z}_\mathrm{PP}\in\mathbb{C}^{Q\times Q}$, respectively, describing the self-impedance for the antenna port and pixel ports; $\mathbf{z}_\mathrm{PA}\in\mathbb{C}^{Q\times 1}$ and $\mathbf{z}_\mathrm{AP}\in\mathbb{C}^{1\times Q}$, respectively, being the trans-impedance relating the voltage of the antenna port to currents of pixel ports, $\mathbf{z}_\mathrm{PA} = \mathbf{z}_\mathrm{AP}^\mathsf{T}$.

Among all $Q+1$ ports in the network, each of the $Q$ pixel ports is further connected to a load impedance $z_{\mathrm{L},q}$ representing the states of a switch, as shown in Fig. \ref{fig:pixel_antenna_model}(b). For on state the load impedance is short-circuit while for off state the load impedance is open-circuit. We describe the on and off states of each switch using a binary variable $b_q\in\{0,1\}$. This allows the binary vector $\mathbf{b} = [b_1,\ldots,b_Q]^\mathsf{T}\in\{0,1\}^{Q\times 1}$, referred to as the \textit{antenna coder}. Therefore, the load impedance matrix is a function of the antenna coder given by $\mathbf{Z}_\mathrm{L}(\mathbf{b}) = \mathsf{diag}(z_{\mathrm{L},1},\ldots,z_{\mathrm{L},Q})\in\mathbb{C}^{Q\times Q}$ with\footnote{The open-circuit state of the switch is numerically approximated as $z_{\mathrm{L},q} = \jmath\gamma$, where $\jmath$ denotes the imaginary unit and $\gamma\in\mathbb{R}^+$ is sufficiently large, such as $\gamma = 10^{10}$.} 
\begin{equation}
    z_{\mathrm{L},q}=\begin{cases}
        \infty, &\text{if } b_q = 1,\\ 
        0, &\text{if } b_q = 0.
    \end{cases}
\end{equation}
$\mathbf{Z}_\mathrm{L}(\mathbf{b})$ relates the voltage $\mathbf{v}_\mathrm{P}$ and current $\mathbf{i}_\mathrm{P}$ by $\mathbf{v}_\mathrm{P} = -\mathbf{Z}_\mathrm{L}(\mathbf{b})\mathbf{i}_\mathrm{P}$. As a result, the currents at the pixel ports can be described as
\begin{equation}
    \mathbf{i}_\mathrm{P}(\mathbf{b}) = - (\mathbf{Z}_\mathrm{PP} + \mathbf{Z}_\mathrm{L}(\mathbf{b}))^{-1}\mathbf{z}_\mathrm{PA}i_\mathrm{A}.
\end{equation}
To derive the radiation pattern of a pixel antenna as a function of the antenna coder $\mathbf{b}$, we first introduce the open-circuit radiation pattern matrix $\mathbf{E}_\mathrm{oc} = [\mathbf{e}_\mathrm{A}^\mathrm{oc},\mathbf{e}_{\mathrm{P},1}^\mathrm{oc},\ldots,\mathbf{e}_{\mathrm{P},Q}^\mathrm{oc}]\in\mathbb{C}^{2K\times (Q+1)}$, which contains the radiation pattern $\mathbf{e}_\mathrm{A}^\mathrm{oc}\in\mathbb{C}^{2K\times 1}$ (comprising $\theta$ and $\phi$ polarization components across $K$ spatial samples) excited by a unit current at the antenna port with all other ports open-circuit, and radiation patterns $\mathbf{e}_{\mathrm{P},q}^\mathrm{oc}\in\mathbb{C}^{2K\times 1}$ $\forall q$ excited by a unit current at the pixel port $q$ with all other ports open-circuit. Given $\mathbf{E}_\mathrm{oc}$, the radiation pattern of the pixel antenna is obtained as the superposition of radiation patterns from all $Q+1$ ports,
\begin{equation}
    \mathbf{e}(\mathbf{b}) = \mathbf{E}_\mathrm{oc}\mathbf{i}(\mathbf{b}),
\end{equation}  
where $\mathbf{i}(\mathbf{b}) = [i_\mathrm{A};\mathbf{i}_\mathrm{p}(\mathbf{b})]$.
By flexibly selecting $\mathbf{b}$ within its feasible range, the radiation pattern $\mathbf{e}(\mathbf{b})$ of the pixel antenna can be reconfigured, thereby providing additional beam manipulation flexibility to improve wireless systems.

\subsection{Formulation for Antenna Coding Optimization}

\begin{figure}
    \centering
    \includegraphics[width=0.45\textwidth]{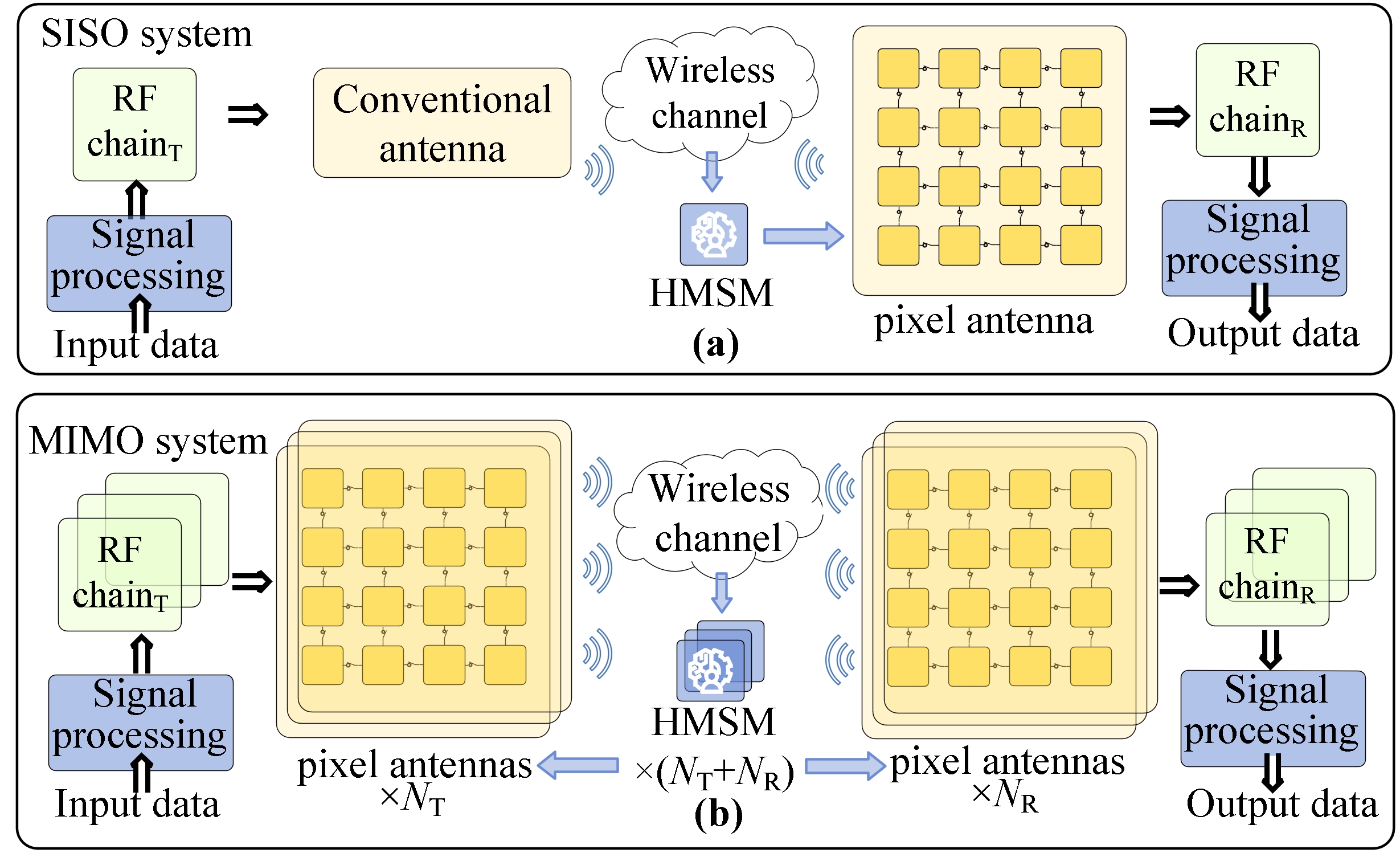}  % 或者使用 width=0.9\textwidth 稍微留边距
    \caption{Diagram of (a) pixel antenna empowered SISO systems and (b) pixel antenna empowered MIMO systems integrated with HMSM.}
    \label{fig:system}
\end{figure}

In this work, we consider two typical use cases of pixel antennas, namely pixel antenna empowered SISO and MIMO systems, as shown in Fig. \ref{fig:system} and explained in the sequel. 

\textit{Case 1: Pixel Antenna Empowered SISO System.} 
A pixel antenna empowered SISO system is first considered where the transmitter is equipped with a conventional antenna with fixed configurations and the receiver is equipped with a pixel antenna, as illustrated in Fig. \ref{fig:system}(a). This is the simplest case utilizing pixel antennas. To explicitly characterize the impact of antenna coding using single pixel antenna on channel gain performance, we consider the beamspace channel expression \cite{4shen2025antenna,han2025exploiting}, i.e. 
\begin{equation}
    h(\mathbf{b}_\mathrm{R}) = \mathbf{e}_\mathrm{R}^\mathsf{T}(\mathbf{b}_\mathrm{R})\mathbf{H}_\mathrm{v}\mathbf{e}_\mathrm{T},
\end{equation} 
where $\mathbf{e}_{\mathrm{R}}(\mathbf{b}_\mathrm{R})$ is the normalized radiation pattern of the pixel antenna at the receiver coded by antenna coder $\mathbf{b}_\mathrm{R}$ with $\|\mathbf{e}_\mathrm{R}(\mathbf{b}_\mathrm{R})\| = 1$ and $\mathbf{e}_{\mathrm{T}}$ is the normalized radiation pattern of the conventional antenna with fixed configuration at the transmitter satisfying $\lVert \mathbf{e}_{\mathrm{T}} \rVert = 1$. Meanwhile, $\mathbf{H}_\mathrm{v}\in\mathbb{C}^{2K\times 2K}$ denotes a virtual channel and can be written as
\begin{equation}
    \mathbf{H}_{\mathrm{v}} = \left[ \begin{array}{ll}
    \mathbf{H}_{\mathrm{v},\theta\theta} & \mathbf{H}_{\mathrm{v},\theta\phi} \\
    \mathbf{H}_{\mathrm{v},\phi\theta} & \mathbf{H}_{\mathrm{v},\phi\phi}
    \end{array} \right],
\end{equation}
where $\mathbf{H}_{\mathrm{v},\theta\theta}$, $\mathbf{H}_{\mathrm{v},\theta\phi}$, $\mathbf{H}_{\mathrm{v},\phi\theta}$, $\mathbf{H}_{\mathrm{v},\phi\phi} \in \mathbb{C}^{K\times K}$ represent the virtual channel matrices for $\theta$ and $\phi$ polarizations, respectively, with each entry denoting the channel gain from an angle of departure (AoD) to an angle of arrival (AoA) across $K$ spatial angle samples\footnote{In this work, we consider a rich scattering environment with Rayleigh fading, assuming $[\mathbf{H}_{\mathrm{v}}]_{i,j} \ \forall i, j$ are independent and identically distributed random variables following a complex Gaussian distribution.}.

Aiming at maximizing the channel gain performance, the antenna coder design problem can be formulated as\footnote{The current at the antenna port of the pixel antenna, denoted as $i_{\mathrm{A}}^\mathrm{R}$, should be a variable to ensure the radiation pattern satisfies that $\|\mathbf{e}_\mathrm{R}(\mathbf{b}_\mathrm{R})\|=1$. However, it can be removed since $i_{\mathrm{A}}^\mathrm{R}$ is essentially a function of $\mathbf{b}_\mathrm{R}$ so that the antenna coder is the only variable in the channel gain maximization problem.}
\begin{equation}
    \begin{aligned}
    \max_{\mathbf{b}_\mathrm{R}} \quad & |\mathbf{e}_{\mathrm{R}}^{\mathsf{T}}(\mathbf{b}_{\mathrm{R}}) \mathbf{H}_{\mathrm{v}}\mathbf{e}_{\mathrm{T}} |^2,\\
    \text{s.t.} ~~~ & [\mathbf{b}_\mathrm{R}]_q \in \{0, 1\}, \forall q. \end{aligned} \label{eq:prob_SISO}
\end{equation}

\textit{Case 2: Pixel Antenna Empowered MIMO System.} 
We next consider a MIMO system, as illustrated in Fig. \ref{fig:system}(b), where the transmitter (receiver) is equipped with $N_\mathrm{T}$ ($N_\mathrm{R}$) pixel antennas, each of which is coded by the antenna coder $\mathbf{b}_{\mathrm{T},n_\mathrm{T}}$ ($\mathbf{b}_{\mathrm{R},n_\mathrm{R}}$). The transmit and receive antenna coders are collected into matrices $\mathbf{B}_{\mathrm{T}} = [\mathbf{b}_{\mathrm{T},1}, \dots, \mathbf{b}_{\mathrm{T},N_{\mathrm{T}}}] \in \mathbb{R}^{Q \times N_{\mathrm{T}}}$ and $\mathbf{B}_{\mathrm{R}} = [\mathbf{b}_{\mathrm{R},1}, \dots, \mathbf{b}_{\mathrm{R},N_{\mathrm{R}}}] \in \mathbb{R}^{Q \times N_{\mathrm{R}}}$.
We formulate the beamspace channel representation for the pixel antenna empowered MIMO system, $\mathbf{H}(\mathbf{B}_\mathrm{T},\mathbf{B}_\mathrm{R})\in\mathbb{C}^{N_\mathrm{R}\times N_\mathrm{T}}$, as
\begin{equation}
\mathbf{H}(\mathbf{B}_{\mathrm{T}}, \mathbf{B}_{\mathrm{R}}) = \mathbf{E}_{\mathrm{R}}^{\mathsf{T}}(\mathbf{B}_{\mathrm{R}}) \mathbf{H}_{\mathrm{v}}\mathbf{E}_{\mathrm{T}}(\mathbf{B}_{\mathrm{T}}),
\end{equation}
where $\mathbf{E}_{\mathrm{T}}(\mathbf{B}_{\mathrm{T}}) = [\mathbf{e}_{\mathrm{T},1}(\mathbf{b}_{\mathrm{T},1}), \dots, \mathbf{e}_{\mathrm{T},N_{\mathrm{T}}}(\mathbf{b}_{\mathrm{T},N_{\mathrm{T}}})] \in \mathbb{C}^{2K \times N_{\mathrm{T}}}$ with $\mathbf{e}_{\mathrm{T},n_{\mathrm{T}}}(\mathbf{b}_{\mathrm{T},n_{\mathrm{T}}})$ being the normalized radiation pattern of the $n_{\mathrm{T}}$th transmit pixel antenna satisfying $\|\mathbf{e}_{\mathrm{T},n_{\mathrm{T}}}(\mathbf{b}_{\mathrm{T},n_{\mathrm{T}}})\|=1$, and $\mathbf{E}_{\mathrm{R}}(\mathbf{B}_{\mathrm{R}}) = [\mathbf{e}_{\mathrm{R},1}(\mathbf{b}_{\mathrm{R},1}), \dots, \mathbf{e}_{\mathrm{R},N_{\mathrm{R}}}(\mathbf{b}_{\mathrm{R},N_{\mathrm{R}}})] \in \mathbb{C}^{2K \times N_{\mathrm{R}}}$ with $\mathbf{e}_{\mathrm{R},n_{\mathrm{R}}}(\mathbf{b}_{\mathrm{R},n_{\mathrm{R}}})$ being the normalized radiation pattern of the $n_{\mathrm{R}}$th receive pixel antenna satisfying $\|\mathbf{e}_{\mathrm{R},n_{\mathrm{R}}}(\mathbf{b}_{\mathrm{R},n_{\mathrm{R}}})\|=1$.

Aiming at maximizing the channel capacity and assuming uniform power allocation at the transmitter, the antenna coder optimization problem can be formulated as
\begin{equation}
    \begin{aligned}
        \max_{\mathbf{B}_{\mathrm{T}}, \mathbf{B}_{\mathrm{R}}} &\log_{2} \left| \mathbf{I} + \frac{P}{\sigma^{2} N_{\mathrm{T}}} \mathbf{H} (\mathbf{B}_{\mathrm{T}}, \mathbf{B}_{\mathrm{R}}) \mathbf{H}^\mathsf{H} (\mathbf{B}_{\mathrm{T}}, \mathbf{B}_{\mathrm{R}}) \right| \\
        \text{s.t.} ~~ & [\mathbf{B}_{\mathrm{T}}]_{i,j} \in \{0, 1\}, \forall i,j, \\
        & [\mathbf{B}_{\mathrm{R}}]_{i,j} \in \{0, 1\}, \forall i,j. 
    \end{aligned}\label{eq:prob_MIMO}
\end{equation}
where $\mathbf{I}$ denotes the identity matrix with a proper dimension, $\sigma^2$ represents the power of additive white Gaussian noise, and $P$ is the transmit power. In this case, we define the signal-to-noise ratio (SNR) as $\frac{P}{\sigma^2}$.

While the above two antenna coding optimization problems (\ref{eq:prob_SISO}) and (\ref{eq:prob_MIMO}), as binary optimization problems, have been solved by some heuristic searching-based algorithms, such as successive exhaustive Boolean optimization (SEBO) \cite{17shen2016successive}, the satisfactory performance is achieved at the cost of high computational complexity. To reduce the computational complexity for antenna coding optimization, in the following section, we will propose a deep learning-based algorithm for both pixel antenna empowered SISO and MIMO systems.

\section{Deep Learning-Based Antenna Coding Design}

\begin{figure*}
    \centering
    \includegraphics[width=0.78\textwidth]{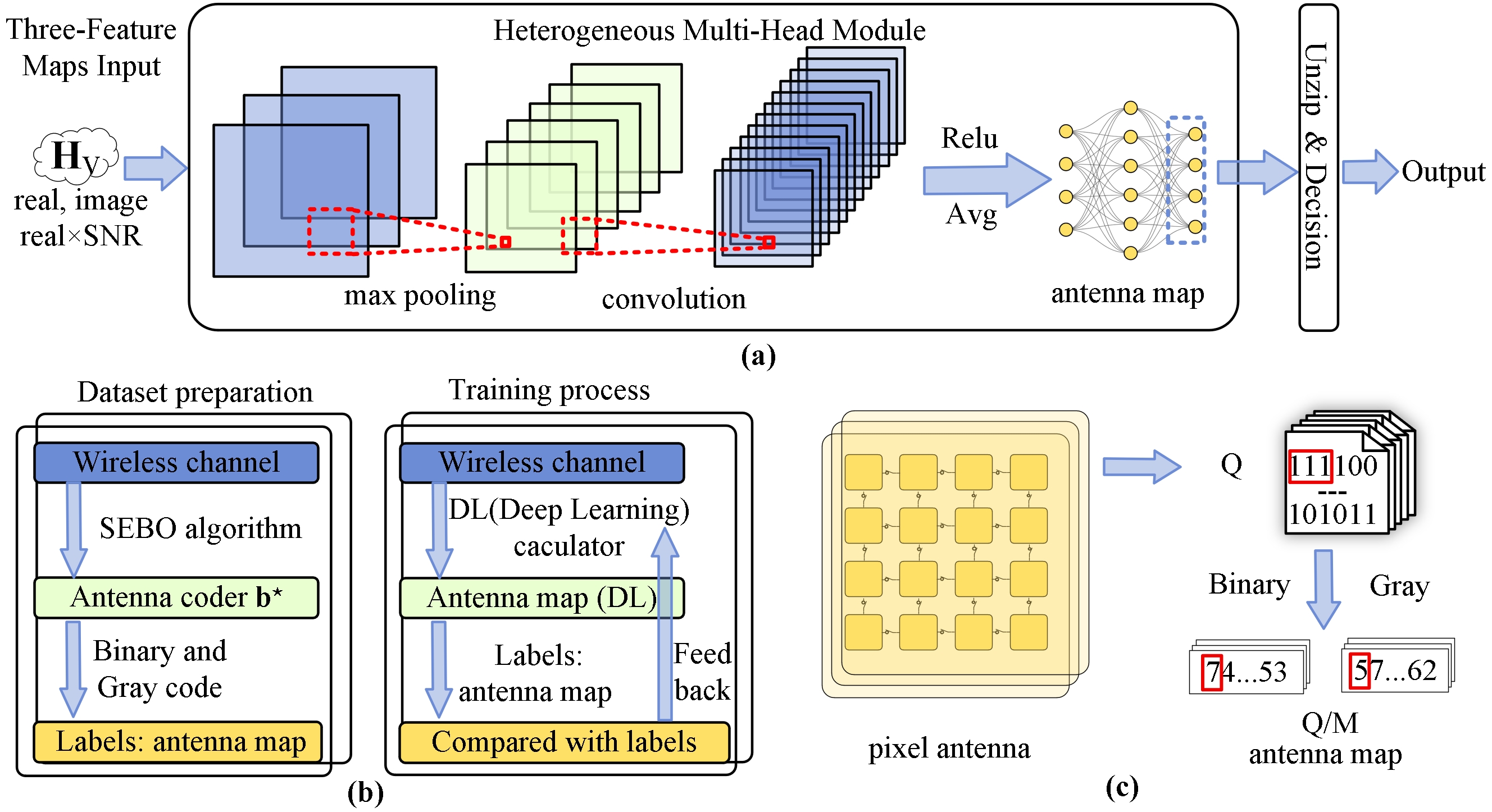}  % 或者使用 width=0.9\textwidth 稍微留边距
    \caption{ (a) Architecture of deep learning-based antenna coding design algorithm. (b) The major components for the model training process, which include the dataset preparation process and the training process. (c) An example of encoding pixel antenna with $Q$ pixel ports into two antenna maps. }
    \label{fig:HMSM}
\end{figure*}

In this section, we propose a deep learning-based antenna coding design algorithm supported by the novel HMSM. 
The proposed HMSM is inspired by the multi-head design whose aim is to enable parallel decoding of the input across multiple independent representation subspaces \cite{15vaswani2017attention}. From a probabilistic perspective, denote the probability that each head correctly produces the output as $p_i$. Then the overall probability of the system accurately producing the correct output given the input, satisfies the following lower bound \cite{19bertsekas2008introduction}
\begin{equation}
P_{\text{correct}} \ge 1 - \left( \prod_{i=1}^{L} (1 - p_i) \right)^{\alpha},
\end{equation}
where $\alpha$ is a correlation-dependent parameter and $L$ denotes the number of heads. This lower bound demonstrates that when the error patterns of different heads are uncorrelated ($\alpha=1$), multi-head collaboration can significantly enhance system robustness, even when the individual accuracy of each head is relatively low.
Below, we will first explain the architecture of HMSM, followed by the whole training process.

\subsection{Architecture of HMSM} 

The proposed HMSM has an architecture comprising three key components: a three-channel input, a heterogeneous multi-head module, and a unzip layer together with a decision function, as shown in Fig. \ref{fig:HMSM}(a). 

\subsubsection{Three-Feature Maps Input} In this work, the three-feature maps are adopted, analogous to the RGB input in computer vision \cite{18koonce2021resnet}. The three-feature maps are constructed as follows: the first two maps are the real and imaginary parts of the virtual channel matrix $\mathbf{H}_{\mathrm{v}}$, i.e., $\Re\{\mathbf{H}_\mathrm{v}\}$ and $\Im\{\mathbf{H}_\mathrm{v}\}$, respectively. In the case of SISO scenarios, the third map is set as $\Re\{\mathbf{H}_\mathrm{v}\}$; in the case of MIMO systems, the third map is formed by the element-wise product of the SNR value and the real part, i.e., $\mathrm{SNR} \times \Re\{\mathbf{H}_\mathrm{v}\}$. This design serves two key purposes: 1) it ensures all input channels have identical dimensions, which is a requirement for standard convolutional neural network architectures; 2) it provides an explicit inductive bias by fusing the SNR information directly with the channel state information. This allows the network to more efficiently learn features that are conditioned on the SNR level, rather than treating SNR as a separate global variable.

\subsubsection{Heterogeneous Multi-Head Module} This component consists of multiple ResNet50 convolutional neural networks \cite{18koonce2021resnet}. Each ResNet50 serves as an individual head, processing the input matrix through pooling and convolution operations. The resulting features are then flattened into vectors and passed through fully connected layers that apply linear transformations for logistic regression. In addtion, each head ultimately outputs the encoded antenna coder (antenna map), the design of which will be detailed in the Training Process section. 

\subsubsection{Unzip Layer and Decision Function} The antenna maps generated by each head are processed through the unzip layer, yielding multiple candidate solutions for the antenna coders $\mathbf{b}_\mathrm{R}$ (or $\mathbf{B}_\mathrm{T}$ and $\mathbf{B}_\mathrm{R}$). The final output $\mathbf{b}_\mathrm{R}^\star$ (or $\mathbf{B}_\mathrm{T}^\star$ and $\mathbf{B}_\mathrm{R}^\star$) is then determined by selecting the optimal candidate through a comparative evaluation using the decision function of $|\mathbf{e}_{\mathrm{R}}^{\mathsf{T}}(\mathbf{b}_{\mathrm{R}}) \mathbf{H}_{\mathrm{v}}\mathbf{e}_{\mathrm{T}}|^2$ for pixel antenna empowered SISO systems or $\log_{2} \left| \mathbf{I} + \frac{P}{\sigma^{2} N_{\mathrm{T}}} \mathbf{H} (\mathbf{B}_{\mathrm{T}}, \mathbf{B}_{\mathrm{R}}) \mathbf{H}^\mathsf{H} (\mathbf{B}_{\mathrm{T}}, \mathbf{B}_{\mathrm{R}}) \right|$ for pixel antenna empowered MIMO systems.

\subsection{Training Process} 

The training process consists of the following steps. 

\textit{Step 1:} The training begins with the preparation of the dataset, as illustrated in Fig. \ref{fig:HMSM}(b). $\Re\{\mathbf{H}_\mathrm{v}\}$, $\Im\{\mathbf{H}_\mathrm{v}\}$, and SNR serve as training data, while the corresponding optimal antenna coder $\mathbf{b}_\mathrm{R}^{\star}$ (or $\mathbf{B}_\mathrm{T}^\star$ and $\mathbf{B}_\mathrm{R}^\star$) is generated by optimizing the antenna coder through the SEBO algorithm. 
    
\textit{Step 2:} The antenna coders are further used to generate antenna maps, referred to as a reformulation of antenna coders via different encoding schemes. These antenna maps serve as different sets of labels for training different heads to achieve a lower $\alpha$, as depicted in Fig. \ref{fig:HMSM}(b). In order to improve the computational accuracy, we employ $L=2$ heads in the considered systems, which necessitates the use of two different encoding schemes. Specifically, every $M$ binary digits in an antenna coder with $Q$ binary entries are encoded into decimal numbers using both binary and Gray coding schemes, producing lower-dimensional antenna maps that serve as training labels as depicted in Fig. \ref{fig:HMSM}(c).

\textit{Step 3:} The deep learning model is trained on an extensive set of labeled samples. The differences between predictions and labels are utilized to provide feedback for model refinement over multiple rounds of training, leading to an accurate computational framework. 

After sufficient training, the well-trained deep learning model can efficiently design high-performance antenna maps that enhance channel gain for pixel antenna empowered SISO systems or capacity for MIMO systems.

\textit{Computational Complexity:}
The computational complexity of the neural network is governed by 1) $f(\frac{Q}{M})$ that represents the model complexity associated with the number of subtasks, i.e., $\frac{Q}{M}$ and 2) $g(2^M)$ that denotes the learning burden introduced by the classification freedom of each subtask. Both are increasing functions. The overall complexity is given by \cite{20andreas2016neural}
\begin{equation}
D(M) = \Phi\left(f\left(\frac{Q}{M}\right), g(2^M)\right),
\end{equation}
where $\Phi(\cdot)$ is a combination function describing how $f(\cdot)$ and $g(\cdot)$ collectively determine the total difficulty. When $M$ keeps increasing while remaining relatively small, the reduction of the number of subtasks $\frac{Q}{M}$ will lead to the reduction of the total complexity. However, when $M$ becomes excessively large, the exponential growth in the number of categories per subtask $2^M$ becomes the dominant factor, leading to increasing overall complexity. Therefore, it is important to properly choose the value of $M$ based on a given $Q$.

\textit{Remark:} The distinction between the two cases lies solely in the label acquisition process. In the SISO case, the labels correspond to the optimal antenna coder $\mathbf{b}^{\star}$ that maximizes the channel gain and are obtained using the SEBO algorithm. In contrast, for the MIMO case, the labels are the optimal antenna coders $\mathbf{B}^{\star}_{\mathrm{T}}$ and $\mathbf{B}^{\star}_{\mathrm{R}}$ that maximize the channel capacity, which are obtained using the SEBO algorithm assuming the transmitter performs a uniform power allocation \cite{4shen2025antenna}.

\section{Performance Evaluation}
We consider a rich scattering propagation environment where each entry in the virtual channel $\mathbf{H}_\mathrm{v}$ are independently and identically complex Gaussian distributed random variables. The angular resolution is set to $\Delta \phi = 5^\circ$, resulting in $K = 72$ angular samples. For the pixel antenna empowered SISO system, the transmit antenna is assumed to have a fixed isotropic radiation pattern, while the receive antenna is a pixel antenna. For the pixel antenna empowered MIMO system, 4 transmit and 2 receive pixel antennas are assumed. Following \cite{4shen2025antenna}, each pixel antenna has a physical aperture of $0.5\lambda\times 0.5\lambda$, where $\lambda=125$ mm denotes the wavelength, and $Q=39$ pixel ports. The impedance matrix $\mathbf{Z}\in\mathbb{C}^{(Q+1)\times(Q+1)}$ and the open-circuit radiation pattern  $\mathbf{E}_\mathrm{oc}\in\mathbb{C}^{2K\times(Q+1)}$ are obtained by the CST studio suite. In the following experiments, each pixel antenna is equipped with an HMSM for efficient and effective antenna coding design. 
We select the number encoded binary digits $M = 3$ to achieve a good balance between the performance and computational complexity.
A dataset of 90,000 unique channel samples is created, with 90\% used for training and 10\% for testing. The SEBO \cite{17shen2016successive} is used with a block size of 12 to maximize the channel gain and capacity and to prepare for the training data.

\begin{figure}
    \centering
    \includegraphics[width=0.8\linewidth]{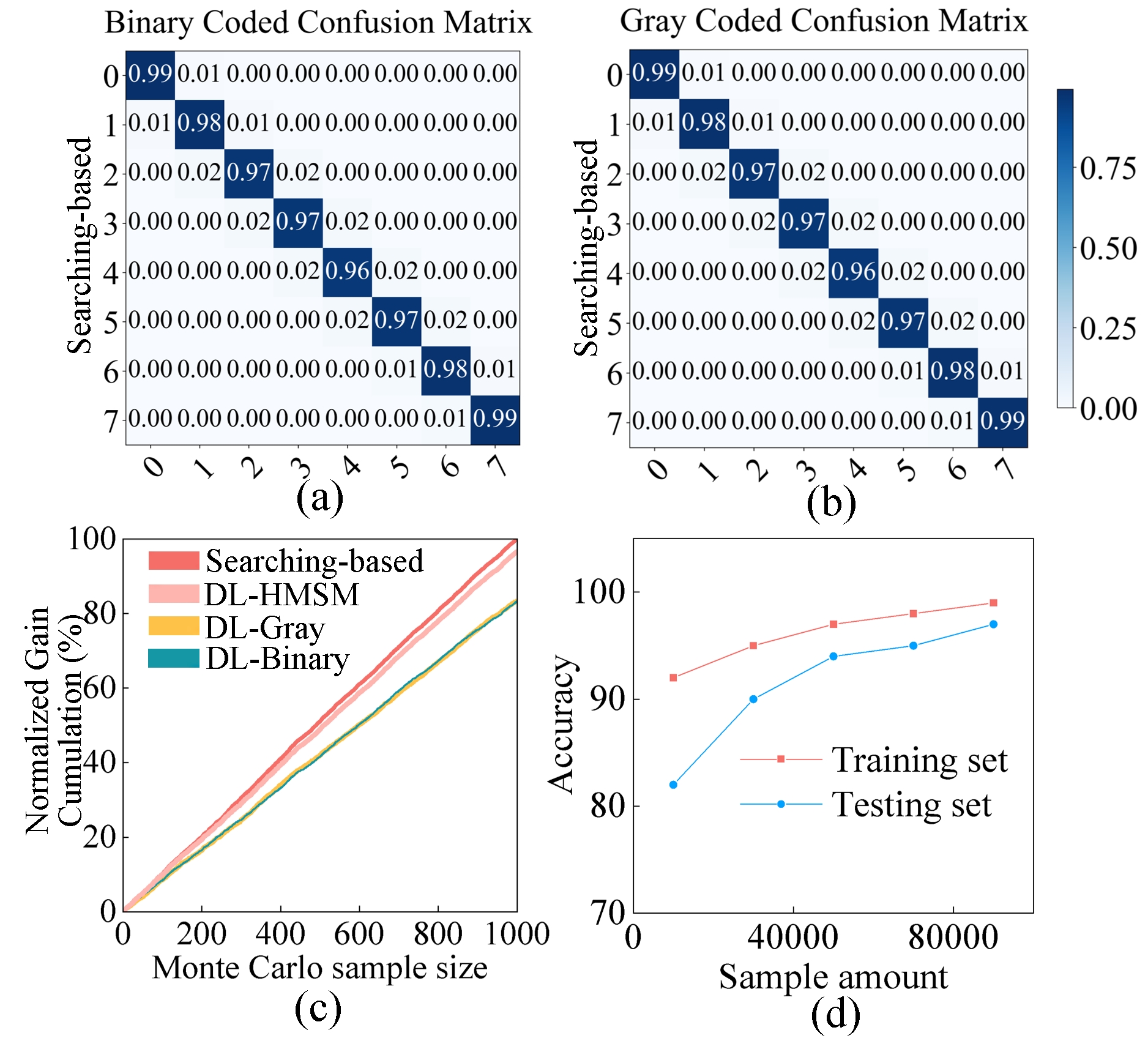}
    \caption{Evaluation of the regression performance of the deep learning model for (a) binary code, (b) Gray code, (c) performance improvement of HMSM, and (d) learning accuracy versus number of samples in the dataset.}
    \label{fig:results_algorithm}\vspace{-0.2 cm}
\end{figure}

Fig. \ref{fig:results_algorithm} evaluates the feasibility and scalability of the deep learning model using either single-head or multi-head mechanisms. Figs. \ref{fig:results_algorithm}(a) and (b) compare deep learning (marked as ``DL'') based antenna maps using binary and Gray coding against the ideal maps obtained by the SEBO algorithm (marked as ``Searching-based''). To quantitatively evaluate the accuracy of a single head, an element-wise\footnote{Each ``element'' refers to one decimal number encoded from $M=3$ binary entries in an antenna coder.} comparison between the antenna map generated by two algorithms is performed. For the case of $M=3$, each element is a discrete category within the set  $\lbrace 0, 1, \ldots, 7 \rbrace$, which is treated as an 8-class classification problem. The performance is analyzed by constructing a confusion matrix, which reveals specific element-wise confusion patterns and prediction accuracy of the trained model. The results show that the neural network effectively captures the nonlinear mapping between $\mathbf{H}_{\mathrm{v}}$ and the antenna coder, achieving an average accuracy of 97.32\%. 
Furthermore, using the SISO gain achieved by the SEBO algorithm as a benchmark, 1000 Monte Carlo samples are performed. Cumulative channel gains are normalized to quantify the performance improvement of the proposed HMSM compared to the single-head scheme using binary or Gray coding. As shown in Fig. 4(c), HMSM significantly improves the performance by selecting the optimal output from multiple heads with different coding schemes, consistent with expectations. In particular, due to the strong nonlinear relationship between the channel and the antenna coder, the performance of the proposed algorithm is highly dependent on the size of the dataset. As illustrated in Fig. 4(d), the model exhibits stronger generalization capability when the dataset approaches the order of $10^5$ samples.

\begin{figure}
    \centering
    \includegraphics[width=0.8\linewidth]{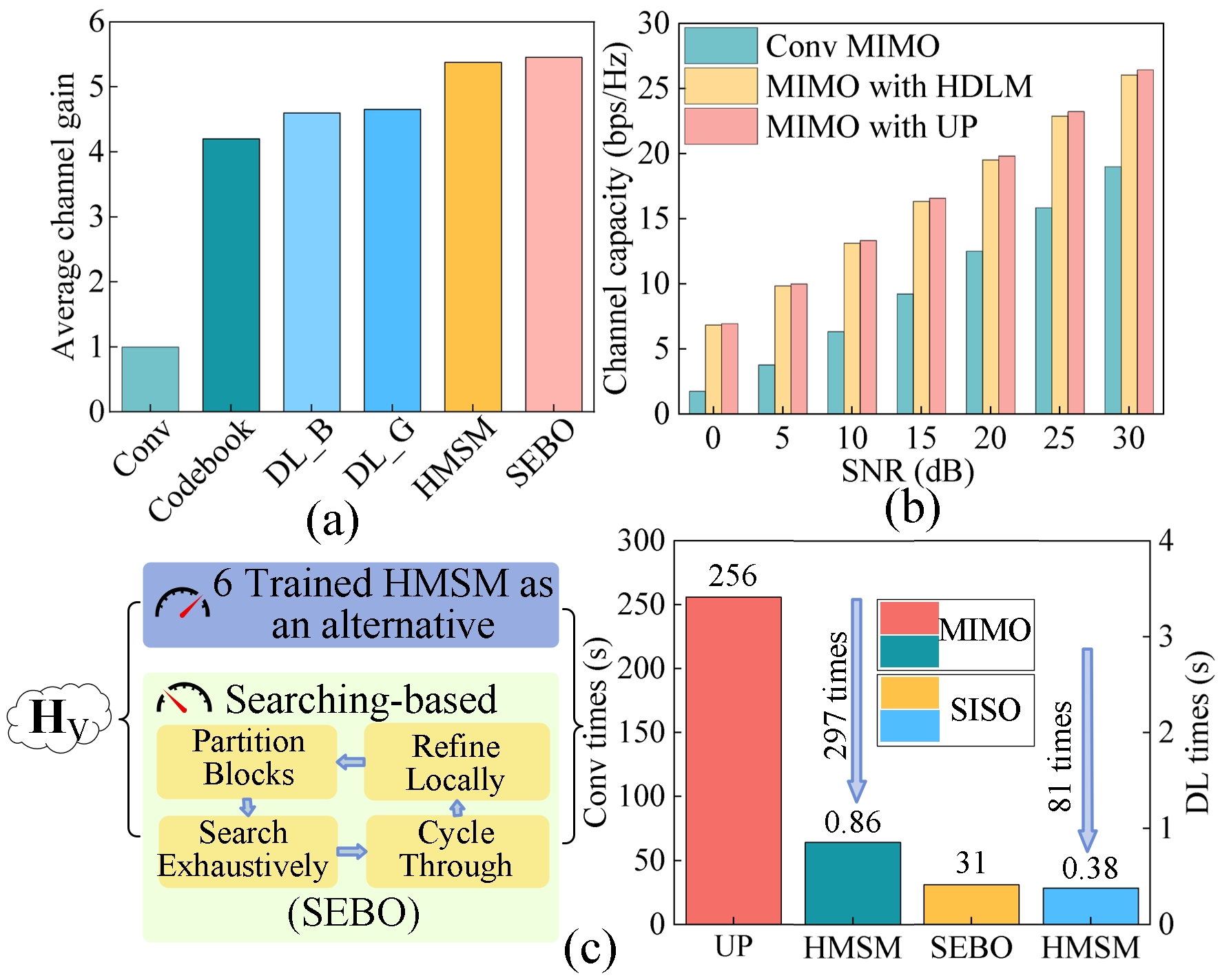}
    \caption{Performance comparison of different algorithms under (a) SISO and (b) MIMO systems, and (c) computational time consumption comparison between the HMSM algorithm and searching-based algorithms.}
    \label{fig:performance}\vspace{-0.2 cm}
\end{figure}

In Fig. \ref{fig:performance}, we evaluate the performance of the proposed HMSM in terms of SISO gain, MIMO capacity, and computational complexity. Fig. 5(a) compares the performance of 1) the single-head deep learning algorithm (marked as ``DL\_B'' and ``DL\_G'' for two coding schemes) and 2) the HMSM-enabled deep learning algorithm with existing methods in the SISO system, including the 1) conventional antenna systems (marked as ``Conv''), 2) codebook-based algorithm \cite{4shen2025antenna} (marked as ``Codebook'' with a codebook size of 1024), 3) and the SEBO \cite{17shen2016successive}. We observe from Figs. \ref{fig:performance}(a) and (c) that, the single-head deep learning algorithms outperform the codebook-based algorithm with much reduced computational complexity.
% the pixel antenna empowered SISO system using the codebook-based algorithms outperforms the conventional antenna system, though with higher computational cost and slightly lower performance than a single-head deep learning-based algorithm. 
The HMSM further improves the average gain by approximately 17\% compared to the single-head deep learning algorithm with slightly higher computational complexity. The performance is only marginally lower than the SEBO algorithm.
In the MIMO model, as described in Fig. 5(b), using HMSM to design the antenna coder also significantly improves channel capacity, achieving around 98.5\% performance of the SEBO algorithm based on uniform power allocation at the transmitter (marked as ``UP''). As shown in Fig. 5(c), 
% replacing heuristic algorithms with neural network methods drastically reduces computation time. 
the time consumption for SISO and MIMO systems using HMSM drastically decreases by $98.77\%$ and $99.66\%$, respectively. In practical experiments, computing multiple models in parallel effectively mitigates the increasing computational time associated with higher computational loads in MIMO systems.

\vspace{-0.1 cm}

\section{Conclusion}
\label{sc:Conclusion}

\vspace{-0.1 cm}

This study develops a deep learning-based antenna coding design algorithm for pixel antenna systems. By utilizing a novel HMSM based on multiple coding schemes, the deep learning-based algorithm achieves satisfactory performance in pixel antenna empowered SISO and MIMO systems with significantly reduced computational complexity. The performance evaluation in SISO systems demonstrates that the proposed algorithm achieves an average channel gain that is of 98.5\% of the searching-based benchmark, while the computational time is only 1.23\% of the searching-based benchmark. In MIMO systems with multiple transceiver antennas, the proposed algorithm achieves an average channel capacity that is 98\% of the searching-based algorithm, while further reducing the computation time to merely 0.34\% of that required by the searching-based algorithm. Both results demonstrate the benefits of the proposed deep learning-based algorithm. 
% Future work will explore increasing the number of heads for a wider variety of wireless channels and more complex environments and reducing the dependency of the model on large training datasets.

\vspace{-0.1 cm}

\section*{Acknowledgment}

The authors would like to acknowledge the support by the Science and Technology Development Fund, Macau SAR (File/Project no.001/2024/SKL), by University of Macau (File no. SRG2025-00060-IOTSC), and by University of Macau Development Foundation (UMDF) (File no. UMDF-TISF-I/2026/025/IOTSC).

\bibliographystyle{IEEEtran}
\bibliography{refs}

\end{document}